\documentstyle[12pt,epsfig] {article}

\begin{document}
\newcommand {\Zzero}     {\mathrm{Z}^0}
\newcommand {\MZ}        {M_{\mathrm{Z}}}
\newcommand {\GZ}        {\Gamma_{\mathrm{Z}}}
\newcommand {\MW}        {M_{\mathrm{W}}}
\newcommand {\GW}        {\Gamma_{\mathrm{W}}}
\newcommand {\GF}        {G_{\mathrm{F}}}
\newcommand {\Gm}        {G_{\mu}}
\newcommand {\thw}       {\theta_{\mathrm{W}}}
\newcommand {\swsq}      {\sin^2\!\thw}
\newcommand {\ee}        {\mathrm{e}^+\mathrm{e}^-}
\newcommand {\WW}      {\mathrm{W}^+\mathrm{W}^-}
\newcommand {\eeWW}    {\ee\to\WW}
\newcommand {\WWg}     {WW\gamma}
\newcommand {\WWZ}     {WWZ}
\newcommand {\eenunug} {\ee\to\nu\bar{\nu}\gamma}
\newcommand {\enuW}    {e^-\bar{\nu}_e \mathrm{W^+}}
\newcommand {\eetoenuW} {\ee\to\enuW}
\newcommand {\eetolnuW} {\ee\to\ell^-\bar{\nu}_{\ell}\mathrm{W}^+}
\newcommand {\eetoenumunu} {\ee\to e^-\bar{\nu}_e\mu^+\nu_{\mu}}
\newcommand {\thetaE}   {\theta_{e^-}}
\newcommand {\costhe}   {\cos\theta}
\def\thefootnote{\fnsymbol{footnote}}
\setcounter{footnote}{1}
\begin{titlepage}

\small
\begin{flushright}
 SAGA-HE-106 \\
 KEK-Preprint 96-83\\
 KEK-CP-048\\
July 1996\\
(Revised: August 1996)
\end{flushright}
\normalsize

\bigskip
\begin{center}
\Large\bf\boldmath 
 Single-W production to test triple gauge boson couplings at LEP
\end{center}
\vspace{1cm}
\begin{center}
\large{Toshio Tsukamoto
\footnote{E-mail address: ttoshio@cc.saga-u.ac.jp}} \\
\normalsize
{\it Department of Physics, Saga University, Saga 840, Japan} \\
\vspace{0.5cm}
\large{Yoshimasa Kurihara} \\
\normalsize
{\it KEK, National Laboratory for High Energy Physics, Ibaraki 305, Japan} \\
\end{center}                                                                    

\vspace{0.5cm}

\begin{abstract}
We present a study of single-W production ($\eetoenuW$)
as a new probe of the anomalous couplings at the LEP energy region.
We introduce simple cuts to separate the single-W process from
W-pair production and have performed cross-section calculations
using 4-fermion generator ``grc4f''.
The cross-section of the single-W process is found to be large enough to
detect at LEP experiments in the near future.
In addition,
a high sensitivity to the anomalous coupling of the $WW\gamma$ vertex
is expected since the amplitude of the $WW\gamma$ diagram makes 
a dominant contribution in this process.
We have found that the cross-section measurement of the single-W process 
in the LEP2 energy region can give complementary bounds on 
the anomalous couplings to those obtained from W-pair analysis.
\end{abstract}
\vspace{1cm}
\begin{center}  
 {\it(To appear in Physics Letters B)}
\end{center}

\end{titlepage}

\section{Introduction}

Recent high precision  measurements performed at LEP and SLC have
clarified that the fermion-gauge boson couplings are amazingly well 
described by the Standard Model (SM).
On the other hand, 
the gauge sector of the Standard Model is still poorly measured.
The non-abelian self couplings of gauge bosons are the most direct
consequences of the $SU(2)\times U(1)$ gauge symmetry and the couplings of
Triple Gauge boson Couplings (TGC) are uniquely determined 
by the Standard Model. Any deviation of these couplings from
their expectation would indicate physics beyond the Standard Model.
While TGCs only enter through loop corrections at LEP1 energy,
a direct confirmation of these couplings can be made at LEP2.

Prior to the actual startup of LEP2,
a number of studies have already been made to find possible ways to 
probe non-standard $\WWg$ and $\WWZ$ couplings~\cite{TGC-genr}. 
Most studies up-to now have focused on the process $\eeWW$.
Although this process is anticipated to give a good sensitivity
to the anomalous couplings, it suffers from the disadvantage that 
one cannot disentangle the effects of $\WWg$ and $\WWZ$ couplings.
Especially, the gauge cancellations between $\gamma$, $\Zzero$ and neutrino
exchange graphs are still not fully operative at LEP2 energy region, hence, 
only the interference effects between different TGCs dominate.
One has to utilise the angular dependence and correlations of the
decay products as much as possible to isolate different linear 
combination of $WWV$ ($V=\gamma, Z$) couplings.

One way to avoid such complications is to use the reaction $\eenunug$ where
only the structure of the $\WWg$ coupling can be studied~\cite{TGC-1gam1,TGC-1gam2}.
The expected sensitivity limits are, however, substantially weaker
at the energy region accessible to LEP~\cite{TGC-1gam1}
because the diagram involved in this channel is WW fusion type.

Another candidate is a single-W process: $\eetoenuW$. 
This process is not well studied so far for LEP.
Pioneering works~\cite{singW-pad} are done to include this `single-W' 
process for the test of TGC. Unfortunately, relatively large polar angle of the
out-going electron is demanded in order to avoid collinear singularities 
in the calculation in ~\cite{singW-pad}. The contribution of real ``single-W'' 
is essentially suppressed as a result. 
In addition, because no separation is made against W-pair production
in~\cite{singW-pad}, the estimated sensitivity is dominated by W-pair
contribution above the WW threshold.
\newpage
In this letter we propose a new probe of the $\WWg$ vertex using
$\eetoenuW$ process where the electron escapes down the beam pipe.
We suggest the selection criteria for single-W production
and present the anticipated sensitivity to the anomalous TGCs.

\section{Single-W production}

\begin{figure}[t]
\begin{center}\hspace*{-2.0cm}
\mbox{
\epsfysize16.cm\epsffile{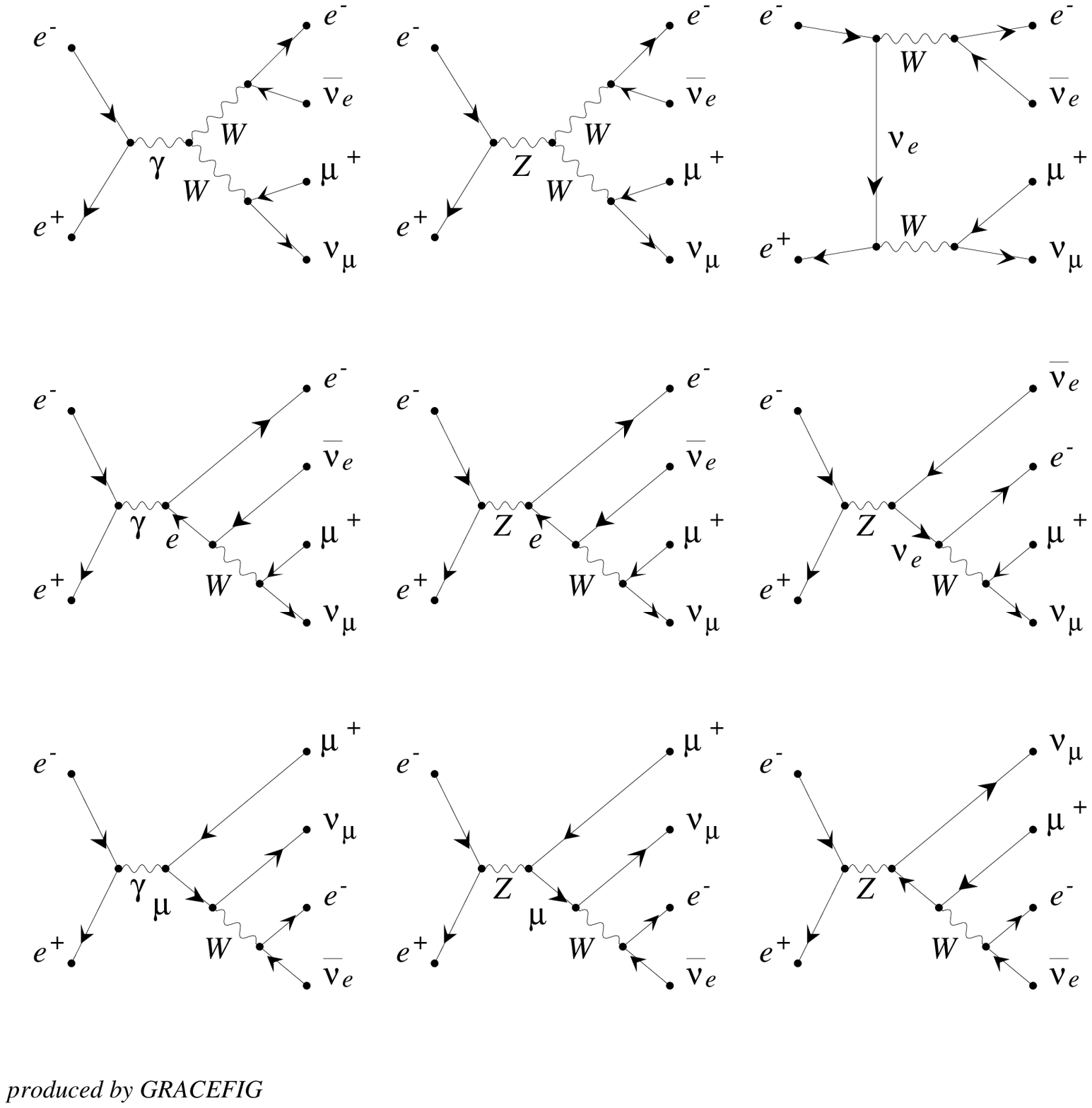}}
\caption{The $s$-channel diagrams of the $\eetoenumunu$ process
in the unitary gauge. The first three diagrams in the first row are
double-resonant diagrams.}
\label{f-diag1}
\end{center}
\end{figure}

\begin{figure}[t]
\begin{center}\hspace*{-2.0cm}
\mbox{
\epsfysize16.cm\epsffile{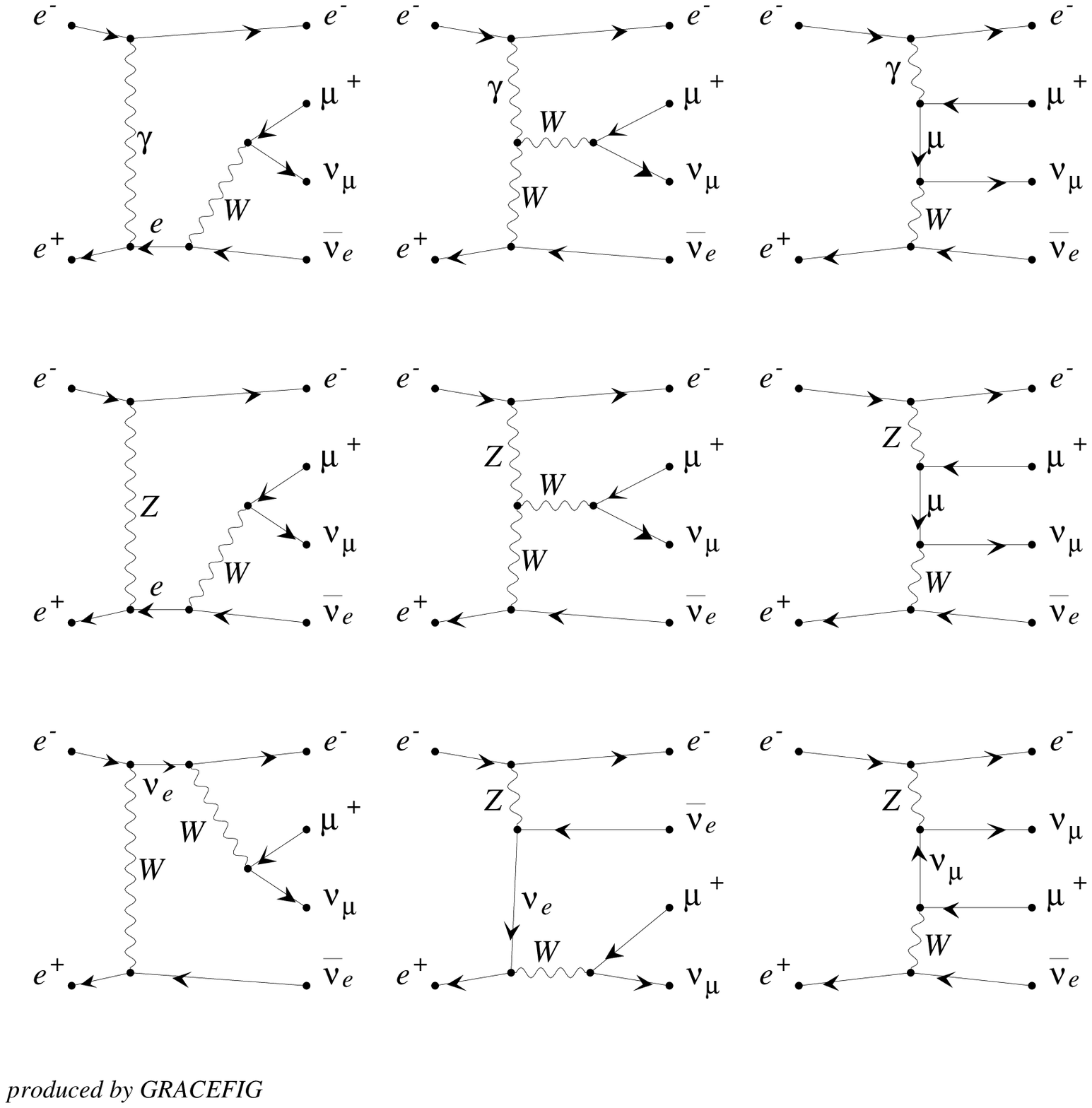}}
\caption{The $t$-channel diagrams of the $\eetoenumunu$ process
in the unitary gauge. The first and second columns show the single-resonant
diagrams and the rest shows the non-resonant diagrams.
Diagrams in the first row ($\gamma$-$W$ processes) gives 
the dominant contribution among the $t$-channel diagrams.}
\label{f-diag2}
\end{center}
\end{figure}

At the energy region of LEP2 and above, four-fermion final state can,
in general, be produced by double (heavy boson) resonant diagrams, 
single resonant diagrams or non-resonant diagrams.
Single-W processes include $\eetolnuW$ where $\l$ can be either
an electron or a muon, and W decays into two fermions.
The electron case is attractive since $t$-channel diagrams also exist.
For example, the diagram involved in $\eetoenumunu$ 
can be grouped into the $s$-channel (figure~\ref{f-diag1})
and $t$-channel classes (figure~\ref{f-diag2}), 
each group forms a gauge invariant set.
Among the $t$-channel diagrams, the $\gamma$-$W$ process 
(the first row in figure~\ref{f-diag2})
give the dominant contribution.
Even below the WW threshold, a significant contribution is expected
from the $\gamma$-$W$-$W$ diagram (the second graph in figure~\ref{f-diag2}).

In order to enhance the $t$-channel contribution, we need to require
the outgoing electron to be within a small angle. This cut causes several
problems for the cross-section calculation.
First of all, the matrix element needs to keep even the electron mass
finite throughout the calculation to avoid a singularity at small
electron polar angle ($\thetaE$). It should be noted that,
although a number of event generators for four-fermion processes
are available now~\cite{MC4fermi}, some approximation, 
including zero mass of fermions, are often introduced 
in order to cope with the complexity of the calculation.
Secondly, the gauge violating term due to the introduction of the finite width
of W is found to blow up at small $\thetaE$~\cite{Gviol-1},
and stops one from obtaining
reliable cross-section as a result. This problem can, however, be
avoided by several methods~\cite{Gviol-2,Gviol-3} 
and the results of each scheme are found to be consistent~\cite{Gviol-3}. 
The third problem is a technical one. Since a large cancellation
between the $\gamma$-$W$ processes 
is expected in the unitary gauge, one has to pay attention to the calculation
of the cross-section especially when a numerical integration is used.
This can be avoided if one takes an efficient gauge.

Here we use ``grc4f''~\cite{grc4f} for the calculation. 
In this package,
all the fermion masses are properly taken into account in the helicity
amplitude (CHANEL~\cite{CHANEL}). A numerical integration 
with multi-dimensional phase space is done by BASES~\cite{BASES}. 
The gauge violation due to the finite width of
the W is cured by subtracting the terms which satisfy the Ward identity,
from the electron current~\cite{Gviol-2}.
This method also benefits the stability of the numerical integration
since each amplitude is safely kept close to unity.
One can thus perform a cross-section calculation reliably even without
a cut on the electron polar angle.
A kinematics suitable for the single-W process is completely
different from that for W-pairs. The package ``grc4f'' incorporates another
set of special kinematics aimed at studying single W processes.
The result presented here have been obtained using the following
set of parameters:
\begin{eqnarray*}
 \MZ &= 91.189 \mathrm{GeV}, \quad \GZ &= 2.497 \mathrm{GeV}, \\
 \MW &= 80.23 \mathrm{GeV},  \quad \GW &= 2.034 \mathrm{GeV},  \\
 \alpha &= 1/128.07,        \quad \swsq &= 1 - (\MW^2/\MZ^2).
\end{eqnarray*}

For the radiative corrections, the electron structure function at
${\cal O}(\alpha^2)$~\cite{RC-SF} is convoluted with the cross-section
for a primary process.
Hard emitted photons with finite transverse momentum may, in general,
distort the direction of the final fermions. 
This effect could not be negligible for the processes with the electron 
(positron) in the final state, for example, the $t$-channel classes of 
$\eetoenumunu$, due to a steep forward-peak of their angular distribution. 
In our approximation, initial state photons are radiated along the beam axis,
and the effect stated above is not taken into account. 
However the global correction of initial state radiation is still useful 
to estimate an experimental sensitivity for TGCs. 
For the future precise measurements of TGCs, more complete treatment 
for the radiative corrections is desired.

\section{Experimental signature}

\begin{figure}[t]
\vspace*{-2cm}
\begin{center}\hspace*{-1.5cm}
\mbox{
\epsfysize17.cm\epsffile{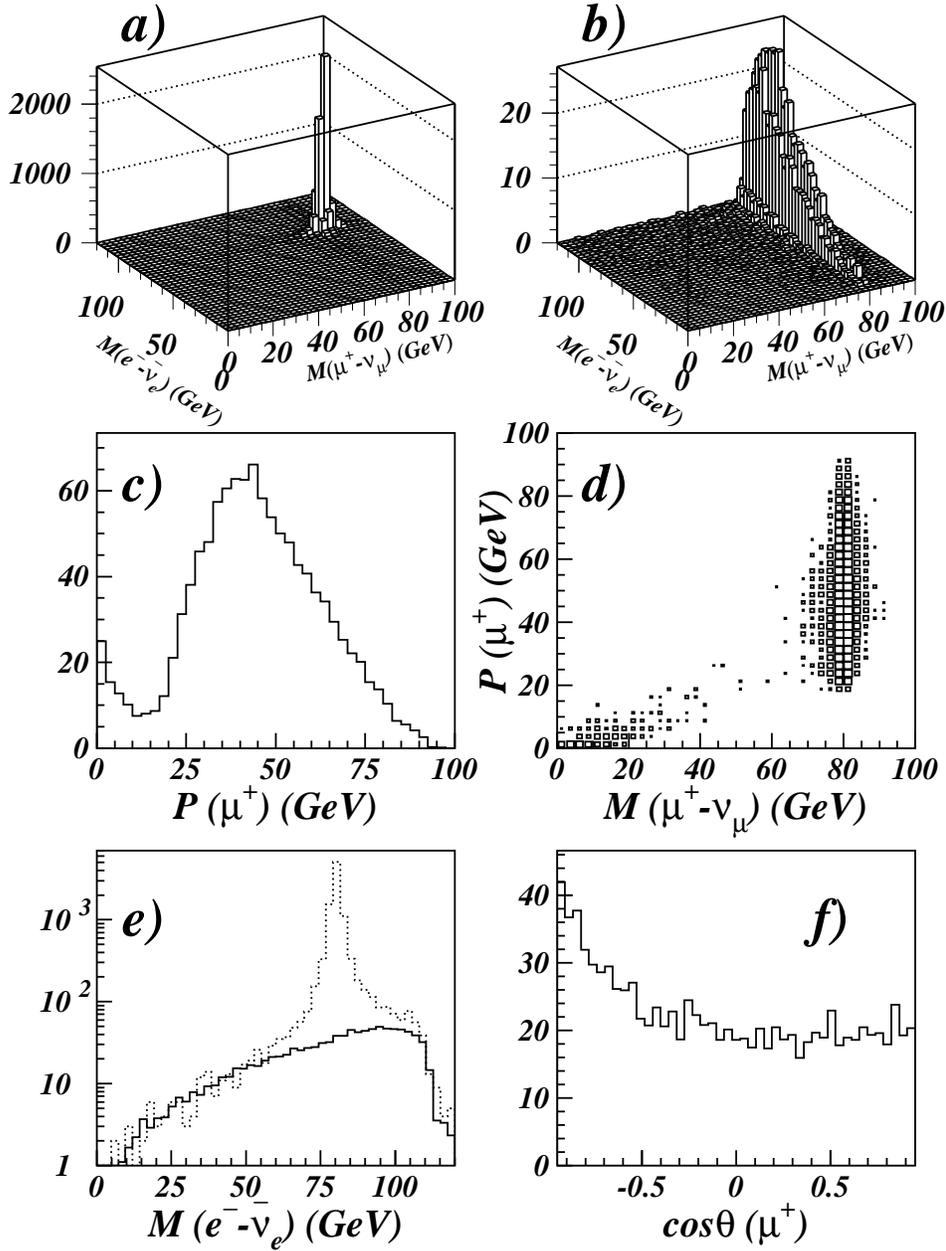}}
\caption{Event distributions for $\eetoenumunu$ process.
The top two figures compares $M(e^{-}$-$\bar{\nu_e})$
versus $M(\mu^+$-$\nu_{\mu}$) with no cuts (a) and 
with the cuts-(\ref{eq:cut1}) (b).
Figure c) and d) show the momentum distribution for muons.
The W-pairs (dashed line in e) are highly suppressed by ``single-W cut'',
while keeping single-W contribution (solid line in e).
The angular distribution for muon is given in f).}
\label{f-evdist}
\end{center}
\end{figure}

We mainly focus on the $\eetoenumunu$ process in this letter 
since the signature is very clean from an experimental point of view.
Once the W-pair threshold opens, one has to discriminate single-W from WW.
This can be done very simply by demanding the electron travels down to
the beam pipe. We use the following cuts on the polar angles
for the electron and the muon.
\begin{eqnarray}
 \thetaE < 35 \,\mathrm{mrad}  \quad\mathrm{and}\quad | \costhe_{\mu^+} | < 0.95.
 \label{eq:cut1}
\end{eqnarray}
The resulting event signature is ``single muon''.
In order to demonstrate how these cuts work, we show several 
event distributions.
The energy and angular distributions for single-W's are not changed much 
at LEP energies 
because it is produced nearly at rest due to the $t$-channel production.
We arbitrarily take the energy point $E_{cm}=$192~GeV.
Figure~\ref{f-evdist}-a) and b) compares the distribution for
the invariant mass of $e^{-}\bar{\nu_e}$: $M(e^{-}$-$\bar{\nu_e})$
versus that of $\mu^+\nu_{\mu}$: $M(\mu^+$-$\nu_{\mu}$) with no cuts 
(\ref{f-evdist}-a) and with the cuts stated above (\ref{f-evdist}-b).
Double resonant contribution is clearly suppressed by the cuts~(\ref{eq:cut1}).

Although the single-W process dominates in the sample after the 
cuts~(\ref{eq:cut1}), a small fraction of extra components exist.
As is seen in figure~\ref{f-evdist}-c) (the momentum of muon: $P_{\mu}$)
and \ref{f-evdist}-d) ($P_{\mu}$ versus $M(\mu^+$-$\nu_{\mu}$)),
this extra contribution is distinct from the single-W production.
Events with low momentum mainly come from the non-resonant diagram
(the third diagram in figure~\ref{f-diag2}) and the rest with high momentum
are from single-W (the first 2 diagrams in figure~\ref{f-diag2}).
With a further requirement of $P_{\mu} >$ 20~GeV, the single-W events can be 
selected. We call this set of cuts ``single-W cuts''.
The $M(e^{-}$-$\bar{\nu_e})$ distribution with ``single-W cuts'' (solid line) 
and with no cuts (dashed line) are compared in figure~\ref{f-evdist}-e),
which illustrates the rejection power against W-pair events.
The angular distribution of the muon after the ``single-W cuts'' are shown
in figure~\ref{f-evdist}-f).
The muon from the single-W process has high energy and its angular
distribution is not particularly forward-peaked. 
Because of this very high Pt signature,
one can remove two-photon events without significant loss of single-W events.

\begin{figure}[t]
\begin{center}\hspace*{-1.0cm}
\mbox{
\epsfysize16.cm\epsffile{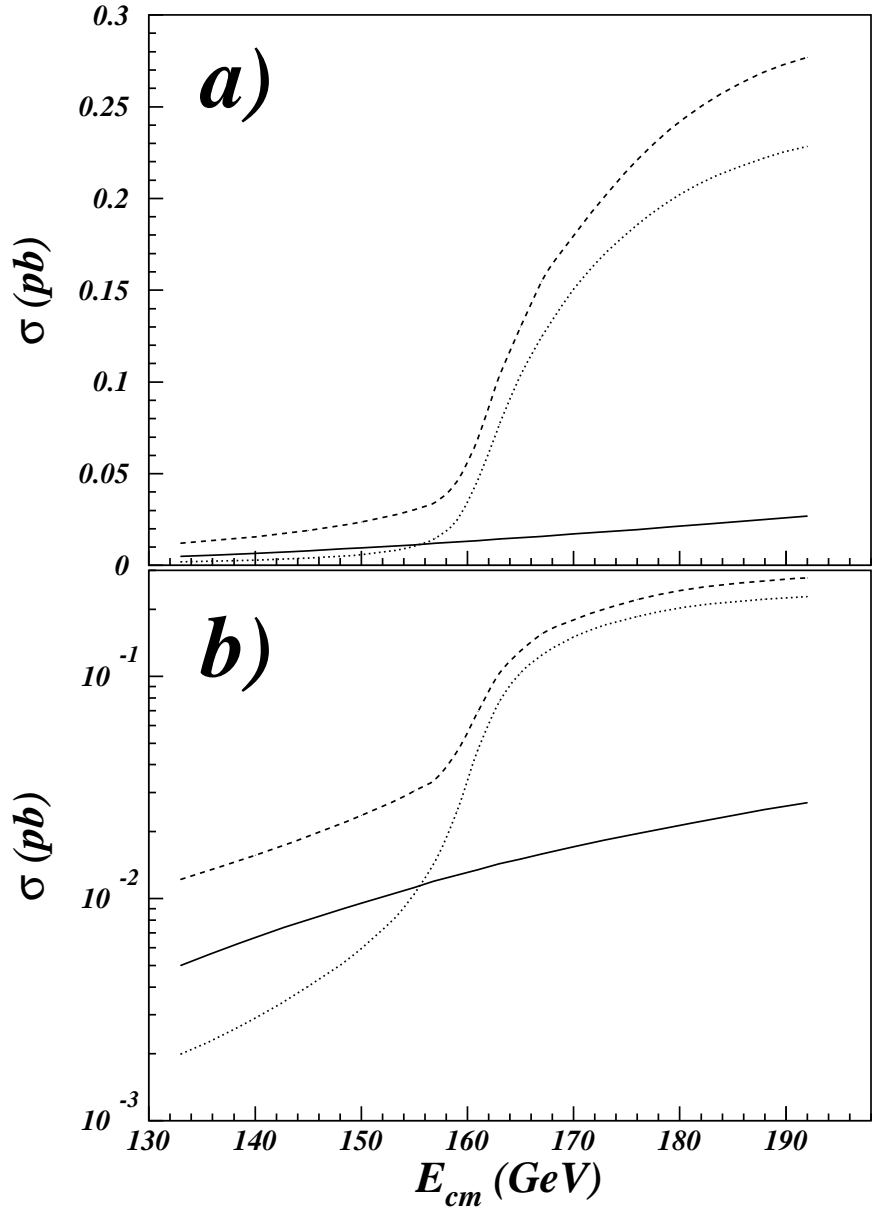}}
\caption{The total cross-section for the process $\eetoenumunu$
as a function of $E_{cm}$ with no cut (dashed line),
with ``canonical cuts'' (dotted line) and with ``single-W cuts'' (solid line)
on a linear scale (a) and a logarithmic scale (b).}
\label{f-xsect}
\end{center}
\end{figure}

We have calculated the total cross-section for  $\eetoenumunu$
in three cases;
\begin{itemize}
\item With no cuts.
\item ``Canonical cuts''~\cite{MC4fermi}: 
  $|\cos\theta_{e^-}|, \, |\cos\theta_{\mu^+}|\!<\!0.985$ 
  and $E_{e^-},  \, E_{\mu^+}\!>\!1$~GeV.
\item ``Single-W cuts'' defined above.
\end{itemize}
Figure~\ref{f-xsect} compares the results as a function of $E_{cm}$.
Since ``canonical cuts'' enhances W-pair production,
the cross-section rises rapidly above the WW threshold.
A monotonic increase in single-W cross-section is seen as $E_{cm}$ increases.
Taking the charge conjugate state into account, the cross-section of the
single-W process with $e\nu\mu\nu$ final state is significant
(table~\ref{tab:xsect}).
One can observe ``single muon'' events from single-W production
with modest luminosity at LEP2.

\begin{table}[h]
\begin{center}
\begin{tabular} {|c|c|c|c|c|}
\hline
$E_{cm}$ (GeV) & 161 & 176 & 188 &  192 \\
\hline
$\sigma$ (fb)  & 28  & 39  & 50  &   54 \\
\hline
\end{tabular}     
\caption[foo]
{\label{tab:xsect}
Cross-section of single-W process with $e\nu\mu\nu$ final state }
\end{center}  
\end{table}

\section{Anomalous couplings}
  To test the triple gauge boson coupling, we use the following effective
Lagrangian~\cite{TGCeffLag} assuming both $C$ and $P$ conservation:
\begin{eqnarray}
 i {\cal L}_{eff}^{WWV} = &g_{WWV} \Bigl[ \,
    g_1^V \left( W^{\dagger}_{\mu\nu}W^{\mu}V^{\nu} 
               - W^{\dagger}_{\mu}V_{\nu}W^{\mu\nu} \right)
    \nonumber \\
 &+ \kappa_V W^{\dagger}_{\mu} W_{\nu} V^{\mu\nu} 
 + \frac{\lambda_V}{m^2_W}  W^{\dagger}_{\rho\nu} W^{\mu}_{\nu} V^{\rho\nu} 
     \,\Bigl]
\end{eqnarray}
\noindent
where $V = \gamma$ or Z, and the overall couplings are
$g_{WW\gamma} = e$, $g_{WWZ} = e \cot \theta_W$, 
$W_{\mu\nu} = \partial_{\mu} W_{\nu} - \partial_{\nu} W_{\mu}$
and $ V_{\mu\nu} = \partial_{\mu} V_{\nu} - \partial_{\nu} V_{\mu}$.

Since $g_1^{\gamma}$ is required to be 1 by electromagnetic gauge
invariance, deviations from the Standard Model are defined as 5 parameters:
\begin{eqnarray}
   \Delta g^Z_1 \equiv (g^Z_1-1), \>\>
   \Delta \kappa_{\gamma} \equiv (\kappa_{\gamma} -1), \>\>
   \Delta \kappa_{Z} \equiv (\kappa_{Z} -1), \>\>
   \lambda_{\gamma}, \>\> \lambda_Z
   \label{eq:tgc1}
\end{eqnarray}

Current best published bounds are obtained by CDF and D$\O$
from studies of $W\gamma$ events~\cite{CDF-D0}:
$-1.6 < \Delta\kappa_{\gamma} < 1.8$,
$-0.6 < \lambda_{\gamma} < 0.6$.

The anomalous TGCs are already severely constrained by low energy 
data~\cite{TGCthLOW}.
The parameters in equation (\ref{eq:tgc1}) are no longer independent
each other in order to protect low energy observables 
from acquiring discrepancies with the experimental data.
Extensive studies by the symmetry requirements have been done 
from this point of view, 
which are excellently summarised in~\cite{TGCthSUM1,TGCthSUM2}.
In case the $SU(2)_L\times U(1)_Y$ gauge symmetry is realized linearly,
only three of the five couplings are found
to be independent~\cite{TGCthLOW,TGCthSUM1,TGCthSUM2} 
by considering the dimension 6 operators
which do not affect the gauge boson propagators at tree level. 
As a result,
the $WWZ$ couplings are related to the $WW\gamma$ ones with the equations:
\begin{eqnarray}
\Delta\kappa_{\gamma} = -\cot^2\theta_W\cdot(\Delta\kappa_Z - \Delta g^Z_1),
 \quad \lambda_{\gamma} = \lambda_Z.
 \label{eq:tgc2}
\end{eqnarray}
The anticipated best sensitivity from the W-pair analysis at LEP2 are 
inferred based on this relation, which are found to be~\cite{TGC-genr}: 
$\Delta\kappa_{\gamma} = 0.06$, $\lambda_{\gamma} = 0.04$ and 
$\Delta g^Z_1 = 0.02$.
Another set of three parameters is relevant for the TGC studies at LEP2.
For example, the set ($\Delta\kappa_\gamma, \Delta\kappa_Z, \Delta g^Z_1$)
with $\lambda_{\gamma} = \lambda_Z = 0$ corresponds to the nonlinear 
realization case with the operators of the lowest 
dimensionality~\cite{TGC-genr,TGCthSUM2}.

\begin{figure}[t]
\vspace*{-1cm}
\begin{center}\hspace*{-1.5cm}
\mbox{
\epsfysize17.cm\epsffile{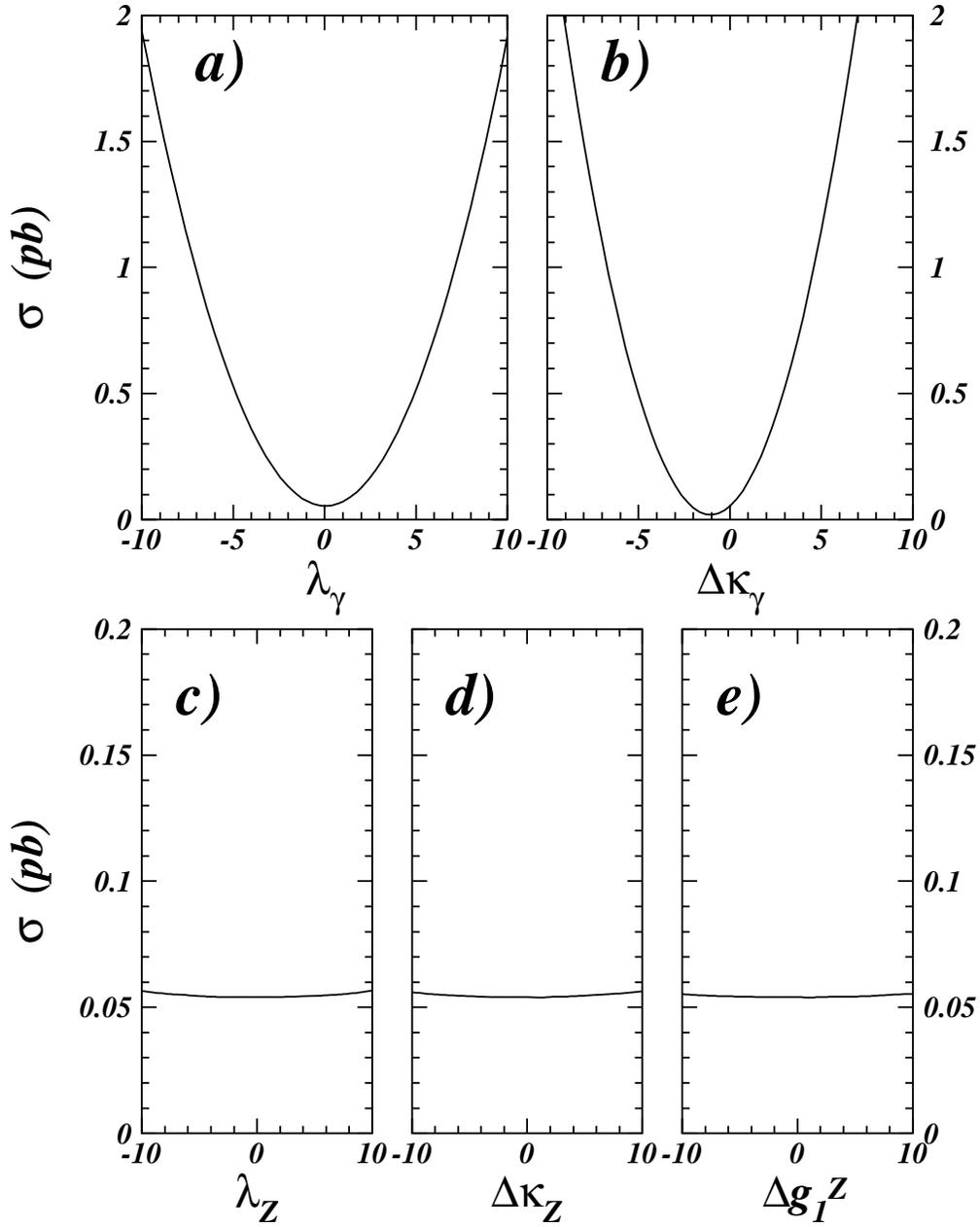}}
\caption{Variation of the cross-section for the single-W process with 
$e\nu\mu\nu$ final state at $E_{cm} = $ 192 GeV:
(a) $\lambda_{\gamma}$, (b) $\Delta\kappa_{\gamma}$,
(c) $\lambda_Z$, (d) $\Delta\kappa_Z$ and (d) $\Delta g_1^Z$.
Note that the vertical scales are expanded for Z-related couplings (c-e).}
\label{f-ACgamz}
\end{center}
\end{figure}

In contrast to the W-pair production, the observable of the single-W
process is not constrained by these relations
since the contribution from the $WWZ$ vertex diagram for single-W process
is very small at LEP energies. In figure~\ref{f-ACgamz},
we demonstrate the variation of the cross-section at $E_{cm}=$ 192~GeV
as a function of the anomalous couplings;
$\lambda_{\gamma}$ (a), $\Delta\kappa_{\gamma}$ (b), 
$\lambda_{Z}$ (c), $\Delta \kappa_{Z}$ (d) and $\Delta g_1^Z$ (e), respectively. 
The cross-section depends only marginally on $WWZ$ related couplings
while we see a large sensitivity to $WW\gamma$ ones.
Because of this orthogonal feature against $WWZ$ couplings,
one can extract anomalous $WW\gamma$ couplings independently by studying
the single-W process even without using any relationship between
$WW\gamma$ and $WWZ$ couplings.

\begin{figure}[t]
\vspace*{-1cm}
\begin{center}\hspace*{-1.5cm}
\mbox{
\epsfysize12.cm\epsffile{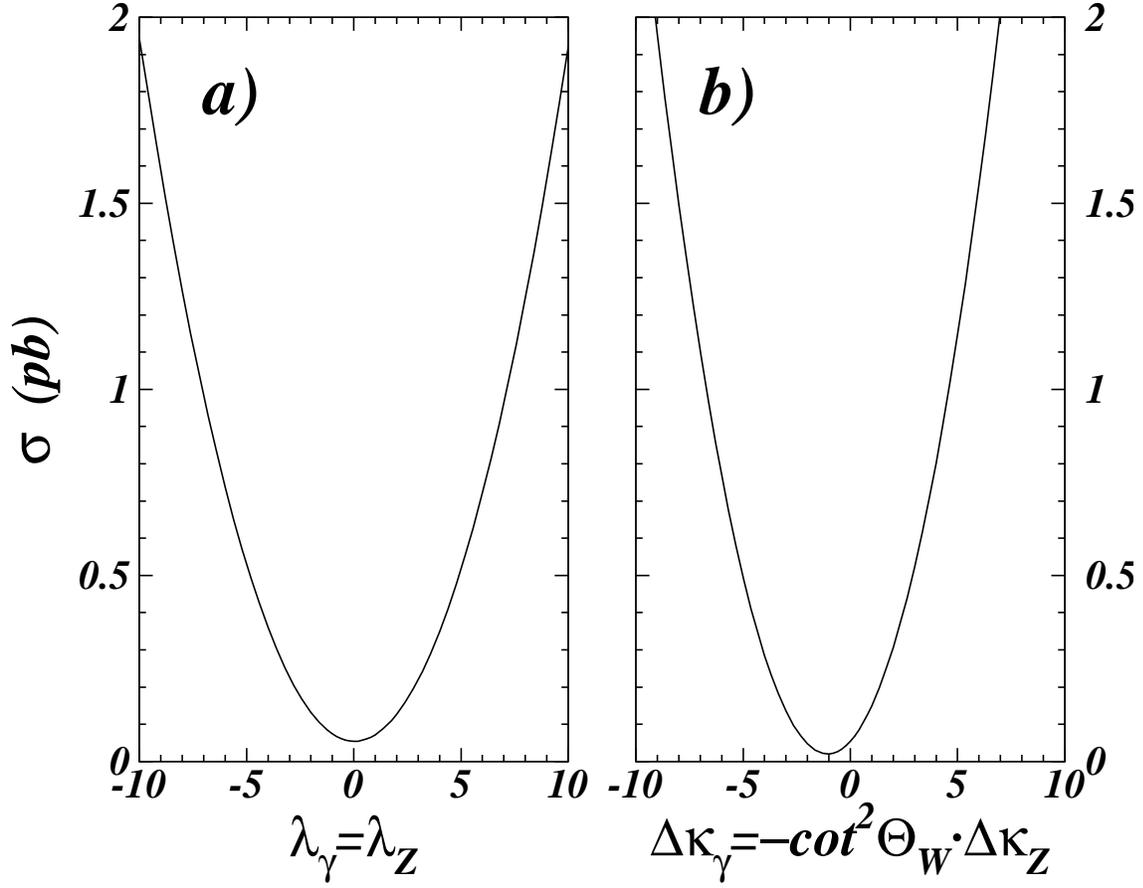}}
\caption{Variation of the cross-section for the single-W process with 
$e\nu\mu\nu$ final state at $E_{cm} = $ 192 GeV with $SU(2)\times U(1)$
constraints:
(a) $\lambda_{\gamma}=\lambda_Z$,
(b) $\Delta\kappa_{\gamma} = -\cot^2\theta_W\cdot\Delta\kappa_Z$.}
\label{f-ACgamz2}
\end{center}
\end{figure}

Figure~\ref{f-ACgamz2} shows the similar sensitivity curves to 
figure~\ref{f-ACgamz} but the constraint~(\ref{eq:tgc2}) is used.
The difference between two cases is numerically very small.
For the rest of this paper, we adopt the 
constraint~(\ref{eq:tgc2}) in order to preserve the gauge invariance.
The third independent quantity $\Delta g_1^Z$ is also insensitive
to the observable and only a huge deviation from the Standard Model
could give rise to a detectable change of the cross-section.
Therefore we approximate it to the Standard Model value (=0).

\begin{figure}[t]
\vspace*{-2.cm}
\begin{center}\hspace*{-1.cm}
\mbox{
\epsfysize16.cm\epsffile{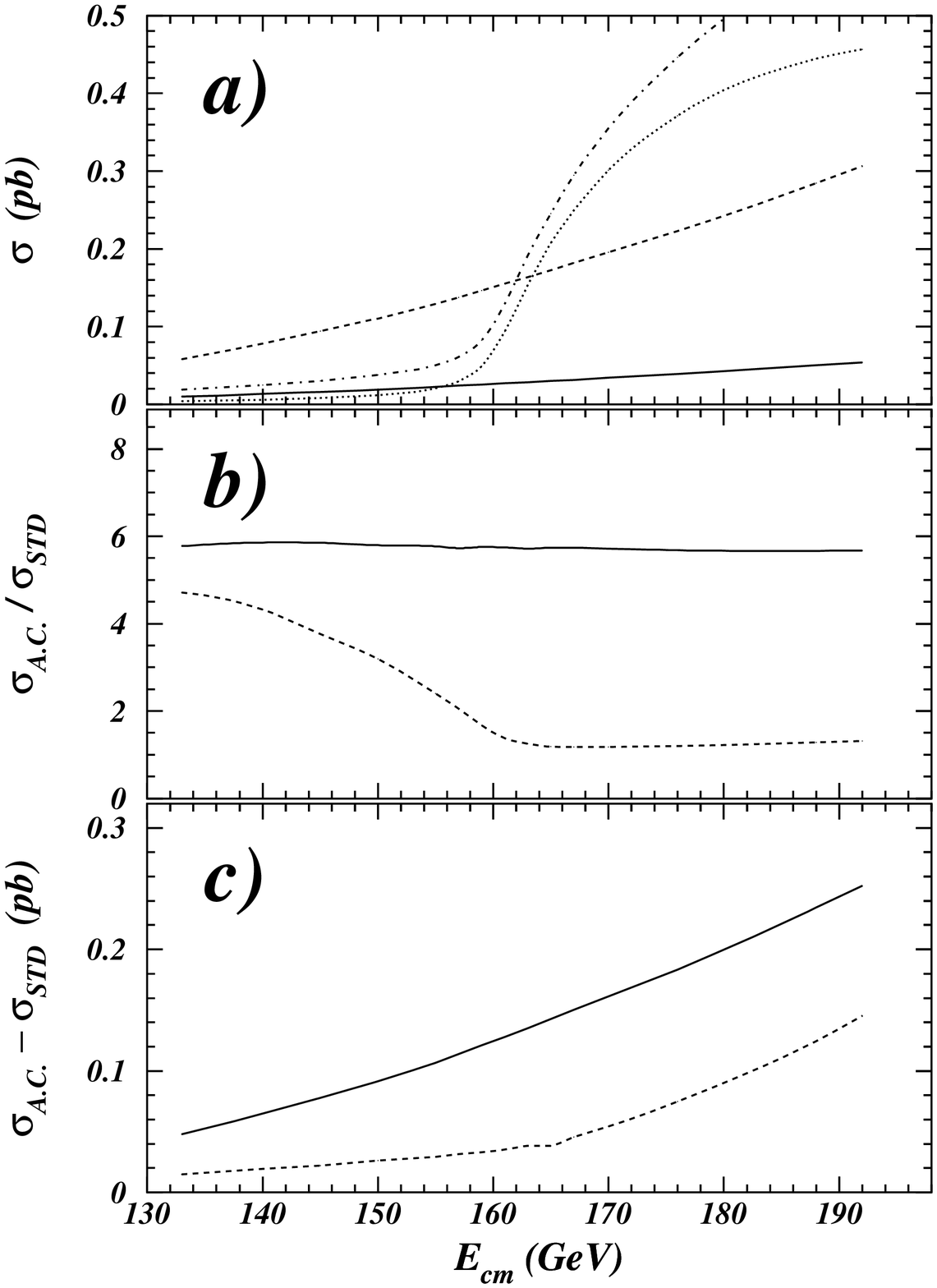}}
\caption{The enhancement of the cross-section due to the anomalous couplings
as a function of $E_{cm}$. 
Two cases are compared: ``single-W cuts'' and ``canonical cuts''.
In the cross-section plot (a), solid (single-W) and dotted (canonical) line  
correspond to the Standard Model cross-section ($\sigma_{STD}$), 
and dashed (single-W) and dashed-dotted (canonical) line  
is with the assumption of 
$\Delta \kappa_{\gamma} = -\cot^2\theta_W\cdot\Delta\kappa_Z= 2$
($\sigma_{A.C.}$).
The ratios of $\sigma_{STD}$ and $\sigma_{A.C.}$ are given (b)
as well as the differences (c) (solid line: single-W cut, dashed-line:
canonical cut).}
\label{f-AC1w_ww}
\end{center}
\end{figure}

Figure~\ref{f-AC1w_ww}-a), b) and c) illustrate a specific 
sensitivity to the anomalous coupling as a function of $E_{cm}$.
In figure~\ref{f-AC1w_ww}-a), the solid line corresponds to
the Standard Model cross-section with ``single-W'' cut.
If we assume anomalous couplings to be 
$\Delta \kappa_{\gamma}  =-\cot^2\theta_W\cdot\Delta\kappa_Z = 2$, 
we get a significantly larger cross-section (dashed line).
Also in the figure, the ratio of two cases (b) and
the difference between two (c) are shown as a function of $E_{cm}$.
The enhancement factor of about 6 is expected,
which is in marked contrast to W-pair production since
the measurement of the WW total cross-section is not sensitive
to the anomalous coupling around the LEP energy region.
We give the sensitivity 
for W-pair production as dashed lines in figure~\ref{f-AC1w_ww}-b) and c), 
where the ``canonical cuts'' are made 
to enhance WW contribution above the threshold.
The W-pair case is much less sensitive than the single-W case.

We estimate the precision in TGCs based on a 1-parameter fit.
The anticipated sensitivity for the anomalous couplings are;
\begin{eqnarray}
 -0.4 < \Delta \kappa_{\gamma} < 0.3, \quad -0.9 < \lambda_{\gamma} < 1.0,
\nonumber
\end{eqnarray}
at 95\% C.L., where $E_{cm}=$ 192~GeV and $\int{\cal L}dt =$ 500pb$^{-1}$ are
assumed.

We emphasise that one can improve the limits further by including
other channels: $e\nu_e e\nu_e$, $e\nu_e \tau\nu_{\tau}$ and 
$e\nu_e q q'$. Event signatures for the other lepton channels are similar to 
single muon case.
The hadronic channel might be difficult to separate from $\gamma\Zzero$
events where the initial $\gamma$ escapes down the beam pipe.
Since a single-W is produced nearly at rest, two jets from the W tend to be
back-to-back and the direction of missing energy does not necessarily lie
close to the beam pipe.
This signature may help for the separation from the $\gamma\Zzero$ process.
In view of TGC studies, the hadronic single-W channel is very attractive
because of its large cross-section.
We have calculated the total cross-section for the hadronic channel
to be 350fb at $E_{cm}=$ 192~GeV.
Single-W process is, in general, more sensitive to $\Delta\kappa_{\gamma}$
than to $\lambda_{\gamma}$. Although the anticipated bound on  
$\lambda_{\gamma}$ will not be attractive ($\sim 0.6$), 
the sensitivity of $|\Delta\kappa_{\gamma}| \sim 0.1$ is expected 
with $\int{\cal L}dt =$ 500pb$^{-1}$ which is comparable to 
that will be obtained from W-pair studies~\cite{TGC-genr}.

\section{Conclusion}
We have for the first time presented the TGC studies
making use of single-W production at the LEP energy region. 
The cross-section measurement of this process 
are found to give good sensitivities to the anomalous couplings,
in particular to $\Delta\kappa_{\gamma}$.

We emphasise that the precise study of the $WW\gamma$ vertex 
from $\eetoenuW$ process is important to disentangle the complex effects
coming from $WW\gamma$ and $WWZ$ vertices
which will be obtained from W-pair production analyses.
In this sense, the bounds from single-W process are complementary
to W-pair ones.

It should also be noted that the sensitivity for the anomalous couplings
does not depend so much on $E_{cm}$ since the cross-section of the single-W 
process is relatively flat over the LEP energy region
while the considerable enhancement is expected with small deviations
from the Standard Model. 
Data collected at any energy points are equally useful.
This signature is in contrast to the W-pair production analysis,
which depends largely on the $E_{cm}$ available. 
Especially even below WW threshold, single-W analysis has a sensitivity
to the anomalous couplings. For example, each LEP experiment
has already collected 5pb$^{-1}$ data at 130-136~GeV. 
Searching for high Pt muons would already
give ${\cal O}$(10) sensitivity to the anomalous coupling constants.

\vspace{0.8cm}
\noindent
{\Large\bf Acknowledgements}
\indent

\vspace{0.5cm}
The authors would like to thank Minami-Tateya theory group of KEK,
especially Y.~Shimizu, J.~Fujimoto and T.~Ishikawa for their 
physics discussions and technical help in grc4f system. 
We also wish to thank T.~Kawamoto for his intuitive discussions
and suggestions at the initial stage of this work.
We are grateful to P.~Watkins for his proofreading the manuscript
which improved the English very much.
This work was supported in part by Japan Society for Promotion of Science
under Fellowship for Research Program at Centres of Excellence Abroad.
\newpage
%

%
\clearpage

%

\begin{thebibliography}{99}

\bibitem{TGC-genr}
G.~Gounaris {\it et al}., `Triple gauge boson couplings' 
in {\it Physics at LEP2}, ed. G.~Altarelli {\it et al}.,
vol.~1, p.525, CERN~96-01, February 1996
and hep-ph/9601233, and references there in.
\bibitem{TGC-1gam1}
G.~Couture and S.~Godfrey, Phys. Rev. {\bf D50} (1994) 5607.
\bibitem{TGC-1gam2}
K.J.~Abraham, J.~Kalinowski, P.\'Saciepko, 
Phys. Lett. {\bf B339} (1994) 136.
\bibitem{singW-pad}
E.N.~Argyres and C.G.~Papadopoulos, Phys. Lett. {\bf B263} (1991) 298; \\
C.G.~Papadopoulos, Phys. Lett. {\bf B333} (1994) 202;\\
C.G.~Papadopoulos, Phys. Lett. {\bf B352} (1995) 144.
\bibitem{MC4fermi}
D.~Bardin {\it et al}., `Event generators for WW physics'
in {\it Physics at LEP2}, ed. G.~Altarelli {\it et al}.,
vol.~2, p.3, CERN~96-01, February 1996.
\bibitem{Gviol-1}
A.~Aeppli, F.~Cuypers and G.J.~van~Oldenborgh, 
Phys. Lett. {\bf B349} (1993) 413; \\
E.E~Boos {\it et al}., Phys. Lett. {\bf B326} (1994) 190.
\bibitem{Gviol-2}
Y.~Kurihara, D.~Perret-Gallix and Y.~Shimizu,
Phys. Lett. {\bf B349} (1995) 367.
\bibitem{Gviol-3}
E.N.~Argyres {\it et al}., Phys. Lett. {\bf B358} (1995) 339; \\
W.~Beenakker {\it et al}., `WW cross-sections and distributions'
in {\it Physics at LEP2}, ed. G.~Altarelli {\it et al}.,
vol.~1, p.79, CERN~96-01, February 1996
and hep-ph/9602351.
\bibitem{grc4f}
J.~Fujimoto {\it et al}., `grc4f v1.1: a Four-fermion Event Generator
for $\ee$ Collisions', preprint KEK-CP-046 and hep-ph/9605312, and
references there in.
\bibitem{CHANEL}
H.~Tanaka, Comput. Phys. Commun. {\bf 58} (1990) 153; \\
H.~Tanaka, T.~Kaneko and Y.~Shimizu, Comput. Phys. Commun.
{\bf 64} (1991) 149.
\bibitem{BASES}
S.~Kawabata, Comput. Phys. Commun. {\bf 41} (1986) 127.
\bibitem{RC-SF}
A.E.~Kuraev, V.S.~Fadin, Sov. J. Nucl. Phys. {\bf 41} (1985) 466.
\bibitem{TGCeffLag}
K.~Hagiwara, K.~Hikasa, R.D.~Peccei, D.~Zeppenfeld,
Nucl. Phys. {\bf B282} (1987) 253; \\
K.~Gaemers and G.~Gouraris, Z. Phys. {\bf C1} (1979) 259.
\bibitem{CDF-D0}
CDF Collaboration, F.~Abe {\it et al}., Phys. Rev. Lett. {\bf 75} (1995) 1017;
D$\O$ Collaboration, S.~Adachi {\it et al}., Phys. Rev. Lett. {\bf 75}
(1995) 1034.
\bibitem{TGCthLOW}
A.De~R\'ujula, M.B.~Gavela, P.~Hern\'andez and E.~Mass\'o,
Nucl. Phys {\bf B384} (12992) 3; \\
K.~Hagiwara, S.~Ishihara, R.Szalapski and D.~Zappenfeld,
Phys. Rev. {\bf D48} (1993) 2182.
\bibitem{TGCthSUM1}
M.~Bilenky, J.L.~Kneur, F.M.~Renard and D.~Schildknecht,
Nucl. Phys. {\bf B409} (1993) 22;\\
H.Aihara {\it et al}., FERMILAB-Pub-95/031.
\bibitem{TGCthSUM2}
F.~Boudjema, Proceedings of the ``Workshop on Physics and Experiments with
Linear e$^+$e$^-$ Colliders'', eds. F.A.~Harris {\it et al}., World Scientific,
1994, p.712.
\end{thebibliography}
\end{document}